\documentstyle[11pt,aaspp4]{article}
\catcode`\@=11
\def\gsim{\ifmmode{\mathrel{\mathpalette\@versim>}}
    \else{$\mathrel{\mathpalette\@versim>}$}\fi}
\def\lsim{\ifmmode{\mathrel{\mathpalette\@versim<}}
    \else{$\mathrel{\mathpalette\@versim<}$}\fi}
\def\@versim#1#2{\lower 2.9truept \vbox{\baselineskip 0pt \lineskip 
    0.5truept \ialign{$\m@th#1\hfil##\hfil$\crcr#2\crcr\sim\crcr}}}
\catcode`\@=12

\def\lsun{\hbox{$L_\odot$}}
\def\msun{\hbox{$M_\odot$}}
\def\nj{\hbox{$N_{\rm j}$}}
\def\bt{\hbox{$B(t)$}}
\def\tj{\hbox{$t_{\rm j}$}}
\def\lt{\hbox{$L_{\rm T}$}}
\def\lb{\hbox{$L_{\rm B}$}}
\def\lv{\hbox{$L_{\rm V}$}}
\def\lk{\hbox{$L_{\rm K}$}}
\def\mbol{\hbox{$M_{\rm bol}$}}

\def\sqr#1#2{{\vcenter{\vbox{\hrule height.#2pt
             \hbox{\vrule width.#2pt height#2pt \kern#2pt
             \vrule width.#2pt}
             \hrule height.#2pt}}}}
\def\square{\mathchoice\sqr34\sqr34\sqr{2.1}3\sqr{1.5}3}
\def\squapp{\hbox{$\square ^{\prime\prime}$}}
\def\boxit#1{\vbox{\hrule\hbox{\vrule\kern3pt\vbox{\kern3pt#1\kern3pt}\kern3pt
             \vrule}\hrule}}
\slugcomment{To appear on the Astronomical Journal}

\lefthead{Renzini}
\righthead{Stellar Populations of Pixels and Frames}

\begin{document}

\title{The Stellar Populations of Pixels and Frames}
\author{Alvio Renzini\altaffilmark{1}}
\affil{European Southern Observatory, Garching b. M\"unchen, D-85748,
Germany}
% Notice that each of these authors has alternate affiliations, which
% are identified by the \altaffilmark after each name.  The actual alternate
% affiliation information is typeset in footnotes at the bottom of the
% first page, and the text itself is specified in \altaffiltext commands.
% There is a separate \altaffiltext for each alternate affiliation
% indicated above.

\altaffiltext{1}{On leave from University of Bologna}

% The abstract environment prints out the receipt and acceptance dates
% if they are relevant for the journal style.  For the aasms style, they
% will print out as horizontal rules for the editorial staff to type
% on, so long as the author does not include \received and \accepted
% commands.  This should not be done, since \received and \accepted dates
% are not known to the author.

\begin{abstract}
Derived from first physical principles, a few simple rules are
presented 
that can  help in both the planning and interpretation of CCD
and IR-array camera observations of resolvable stellar populations.
These rules concern the overall size of the population sampled by a
frame as measured by its total luminosity, and allow to estimate the
number of stars (in all evolutionary stages) that are included in the frame. 
The total luminosity sampled by each pixel (or resolution element) 
allows instead to estimate to which depth meaningful stellar photometry 
can be safely attempted, and below which crowding makes it impossible.
Simple relations give also the number of pixels (resolution elements)
in the frame that will contain an unresolved blend of two stars of any
kind. It is shown that the number of such blends increases
quadratically with both the surface brightness of the target, as well
as with the angular size of the pixel (or resolution element).
A series of examples are presented illustrating how the rules are
practically used in concrete observational situations.
Application of these tools to existing photometric data for the inner
parts of 
the bulge of M31, M32 and NGC 147 indicates that no solid evidence has
yet emerged for the presence of a significant intermediate age population in
these objects.
\end{abstract}

% The different journals have different requirements for keywords.  The
% keywords.apj file, found on aas.org in the pubs/aastex-misc directory, 
% contains a list of keywords used with the ApJ and Letters.  These are 
% usually assigned by the editor, but authors may include them in their 
% manuscripts if they wish. 


% That's it for the front matter.  On to the main body of the paper.
% We'll only put in tutorial remarks at the beginning of each section
% so you can see entire sections together.

% In the first two sections, you should notice the use of the LaTeX \cite
% command to identify citations.  The citations are tied to the
% reference list via symbolic KEYs.  We have chosen the first three
% characters of the first author's name plus the last two numeral of the
% year of publication.  The corresponding reference has a \bibitem
% command in the reference list below.
%
% Please see the AASTeX manual for a more complete discussion on how to make
% \cite-\bibitem work for you.   

\section{Introduction}

The use of sophisticated  software packages for the photometry of stars in
crowded
fields is now a widespread activity. Routines take a CCD frame,
and automatically flatfield it, subtract dark, bias and sky, remove cosmic ray
events, restore bad pixels, and finally deliver a catalog of star
magnitudes and positions. To some extent, what is left to the astronomer is to 
plot color-magnitude diagrams, sort out the astrophysical
inperpretation of them, and finally write the paper.
The use of these packages has thus resulted in a tremendous
progress in the study of resolved stellar populations, including those
in galactic and extragalactic globular clusters, the Magellanic
Clouds, the bulges of the Milky Way and M31, the dwarf satellites of
M31, and several irregular galaxies in the outskirts of the 
Local Group and beyond.
Such progress has been possible thanks to the  large size of the
analyzed samples of stellar populations, coupled with a superior
photometric accuracy. No doubt such achievements would have been
impossible with the old, traditional method of direct optical
inspection of each stellar image. However, relying entirely on the
automatism of the procedures can also produce nonsense, and
occasionally it really did so.

In this paper it is argued that a straighforward application of some 
simple, yet basic  physical concepts can greatly help the
astronomer's work on stellar populations. 
 Some simple tools
will be presented that allow one to easily get the required numbers
out for both the planning for a most
efficient use of the available telescope facilities,
as well as for achieving a deeper
understanding of the scientific meaning of the data
once they have been obtained. 

The paper is organized as follows. Section 2 presents a few basic
conceptual tools whose application to CCD photometry of stellar
populations is then illustrated in the following sections. Section 3
deals with the overall size of a sampled stellar population, hence
with the stellar evolutionary phases that can be investigated given
such size. Section 3 also deals with the size of the stellar
population that is sampled by each resolution element, hence with the
depths (limiting magnitude) that can be achieved before crowding
effects hamper meaningful stellar photometry. In Section 4 several
 examples are presented to illustrate the practical use of such tools,
and Section 5 summarizes the main conclusions of this paper. 

\section{The Number-Luminosity Connection}

In a stellar population of given age the number $\nj$ of stars in any
individual post main sequence (PMS) evolutionary stage is clearly
proportional to the total bolometric luminosity of the population
($\lt$) and
to the duration of the stage ($\tj$), i.e.,:
\begin{equation}
\nj=\bt\lt\tj.
\end{equation}
The coefficient of proportionality $\bt$ is the {\it specific 
evolutionary flux} of the population (Renzini 1981, 1994; Renzini \&
Buzzoni 1986; Renzini \& Fusi Pecci 1988). It is the number of star
entering or leaving any PMS evolutionary stage per year and per solar
luminosity of the population. The product $b(t)=\bt\lt$ then gives the
global rate at which stars leave the MS (in stars per yr), which is
very close to the rate at which stars enter or leave any subsequent
evolutionary stage. Thus, $b(t)$ is also a good approximation the
stellar death rate of the population.

The specific evolutionary flux is a very
weak function of age, ranging from $\sim 0.5\times 10^{-11}$ to
$\sim 2.2\times 10^{-11}\; {\rm stars}/\lsun /{\rm yr}$, as age
increases from $10^7$ yr to 15 Gyr, as illustrated in Fig. 1.
It is also worth noting  that the specific evolutionary flux is almost
independent of the initial mass function (IMF) of the population.
This is because at any age the stars that contribute most of the light
span only a very narrow range of masses, around the mass at the MS turnoff,
and PMS stars span an even narrower range (cf. Fig. 1.1 in Renzini, 1994)

\placefigure{fig1}

Of course, this is not the case for the number of stars still on the
MS,
which is very sensitive to the IMF, $\psi(M)=AM^{-(1+x)}$,
especially towards the lower mass limit.
The IMF scale factor $A$ establishes the size of the population, hence
$A\propto\lt$. With simple algebra one can show that:
\begin{equation}
A=\bt\lt M_{\rm TO}^{1+x}|\dot M_{\rm TO}|^{-1} 
\end{equation}
(Renzini 1994), where $M_{\rm TO}(t)$ is the mass of stars at the MS
turnoff at age $t$. Fig. 2 shows $A/\lt$ as a function of age for
a population with solar composition. Note that for a  population 15
Gyr old $A\simeq 1.2\lt$, and the dependence on the IMF is fairly
weak. For such a population the number of MS stars in a given mass
interval is then given by 
\begin{equation}
dN\simeq 1.2\lt M^{-(1+x)}dM.
\end{equation}
Finally, one may wish to use the luminosity in a given band rather
than the bolometric luminosity, because such a luminosity can be more
directly derived from the observations. To this end one needs {\it
bolometric corrections} for the stellar population, i.e., the
quantities $BC_\lambda$ such that $\lt =BC_\lambda L_\lambda$.
These quantities can be derived theoretically from population
synthesis models, or empirically via multiband observations. 
For the reader's convenience 
Fig. 2 shows the run with age of such bolometric corrections for
three widely used bands (i.e., $B$, $V$, and $K$) from the population
synthesis models by Maraston (1997). For a solar metallicity,
15 Gyr old population one has
\begin{equation}
\lt\simeq  2.2\lb\simeq 1.6\lv\simeq 0.36\lk
\end{equation}
which allow to replace $\lt$ in equation (1)-(3). The same
coefficients in
the models of Buzzoni (1989) have the values 2.5, 2.1, and 0.43, respectively.
In summary, equation (1) allows to estimate the number of stars in
each PMS phase, while an integration of the IMF with the scale factor
given by equation (2) gives the number of MS stars.  
\placefigure{fig2}

\section{The Population Sampling}

Several very useful consequences follow from the simple fact that
 the number of stars in a given phase is
proportional to the sampled luminosity. A
straightforward application of equations (1)-(3) can indeed tell 
a great deal
about the astrophysical information that can be extracted from a given
observation.

\subsection{The Stellar Population of a CCD Frame}

There are two ways to estimate the sampled luminosity. One, especially
useful for planning the observations, relies on surface brightness
information available in the literature. 

Suppose the  CCD camera has a field
of view of $F_{\rm oV}\;\squapp$, and let the surface brightness,
extinction, and true distance modulus of the object be
$\mu_\lambda$ mag/$\squapp$, $A_\lambda$ mag, and ``mod'',
respectively. The sampled luminosity is therefore:

\begin{equation}
\lt=BC_\lambda F_{\rm oV}\, 10^{-0.4(\mu_\lambda -A_\lambda -{\rm
mod}-M_{\lambda\odot})},
\end{equation} 
where $M_{\lambda\odot}$ is the absolute magnitude of the sun in the
$\lambda$ band.

The second method, most
appropriate after the data have been obtained, makes use of the data
themselves. Once the CCD frame has been dark, bias and sky subtracted,
flatfielded, calibrated, and cosmic rays have been removed, then the
total number of counts in the whole frame (the sum of the pixel values
of all the pixels) can be immediately translated into an apparent
integrated magnitude of the whole frame. Hence, applying extinction
and distance corrections one gets the total luminosity sampled by the
frame in a given band. The only steps that require some care are sky
subtraction and saturated images. Obvious precautions include
measuring the sky well outside the object, removing bright interlopers,
and using shallow frames without saturated images. Getting the sampled
luminosity in this way should be done routinely, for this is the most
correct and straightford way to estimate $\lt$, and this quantity is
of great utility for the science to be done with the data.

\placetable{tbl1}

For the sake of concreteness, let us  suppose that the
camera samples a luminosity $\lt=10^5\lsun$, and that the stellar
population is $\sim 15$ Gyr old and of near solar metallicity. 
Table 1 then gives the
number of stars in each of a series of representative evolutionary phases. 
For the MS the number of stars is obtained integrating the IMF from
the
hydrogen burning limit at $M_{\rm inf}=0.1\msun$ to the MS turnoff at $M_{\rm
TO}=0.9\msun$,
and using equation (3). For example, assuming a single slope Salpeter IMF:
\begin{equation}
N_{\rm MS}=1.2\lt\int_{0.1}^{0.9}M^{-2.35}dM\simeq 19.(\lt/\lsun).
\end{equation}
The other entries in Table 1 include: the subgiant branch (SGB), with
stars from the MS turnoff to the
base of the red giant branch (RGB): the RGB itself which ends with helium
ignition at the RGB tip; RGB stars within one magnitude from the RGB
tip (RGBT); the horizontal branch (HB); the early asymptotic giant branch
(E-AGB); the thermally pulsing AGB (TP-AGB); the long period variable
phase (LPV);
the Post-AGB phase (P-AGB), from the AGB tip down to a luminosity ten
times lower; the planetary nebula phase (PN); the white dwarf phase,
from the end of the P-AGB phase down to a luminosity $\sim 10^{-3}\lsun$;
the blue stragglers (BS); and finally the TP-AGB progeny of BSs (BS-TP-AGB).
The LPVs are likely to be in the TP-AGB phase. They are not found in
metal poor globular clusters of our Galaxy; they are found instead
in the old metal rich globulars that belong to the Galactic bulge,
where
they can reach $M_{\rm bol}\simeq -5.0$ (Frogel \& Elias 1988;
Guarnieri, Renzini \& Ortolani 1997). It is worth emphasizing that
only objects brighter than this limit could be indicative of an
intermediate age population.
Lifetimes in the various phases are derived from tested evolutionary
models (e.g., Renzini \& Fusi Pecci 1988). Equation (1) is used for
all phases except for MS, TP-AGB, BS, and BS-TP-AGB. For the MS phase
equation (6) is used. The $10^6$ yr lifetime for the TP-AGB phse is
likely
to be an overestimate for a 15 Gyr old population (cf. Renzini \& Fusi
Pecci 1988). BSs are a trace population of merged binaries, and their
number/luminosity conversion must be  established empirically. For
this purpose the globular
cluster M3 was used (Renzini \& Fusi Pecci 1988), while for the BS-TP-AGB
phase the recipe given by Renzini \& Greggio (1990) was adopted.

Table 1 illustrates that for the given size of the population
($10^5\lsun$) several
evolutionary phases are well represented, while for others the number
of stars is very scarce. When such a number is smaller that unity it can
be used as the probability to find a star of a given kind in the frame. 
For example, with $\lt=10^5\lsun$ the probability of finding a
planetary nebula in the frame is only $\sim 0.5\%$, hence a luminosity
$\sim 2\times 10^7\lsun$ should be explored in order to have a reasonable
chance to find one PN, in practice, the whole globular cluster family
of a galaxy like the Milky Way.

Therefore, Table 1 together with the simple relations given above can
be used to tailor the observations in order to collect the appropriate
number of
stars in particular evolutionary phases, depending on the science goals
of any specific project. Indeed, for any kind of stellar systems one
can chose to point the telescope at a region where the surface
brightness is appropriate, hence the  luminosity sampled by the camera
has the adequate size for the specific science project. Equation (1)
can also be used in the reverse direction, i.e. to infer the duration
of a specific evolutionary phase from the observed number of stars in
the phase and from the sampled luminosity. For example, the duration
of the LPV phase in metal rich globular clusters was derived in this way
(Renzini 1993).

\subsection{The Stellar Population of a Pixel}

To sample an adequate number of stars one may be tempted to observe a
field in
the central regions of an object, where the surface brightness and the
sampled luminosity are highest. However, by moving to high surface
brightness regions crowding will inevitably degrade the photometric
accuracy, and meaningful stellar
photometry may  even become impossible. Clearly, an optimization is
required finding a compromise  between the
conflicting need of securing an adequate size for the sampled
population, and the need for accurate photometry of the stars to be
measured. The effects of crowding will depend on the specific code that
is adopted, with some codes being better than others. However, a great
deal about the effects of crowding can be understood {\it from first 
principles}. 

Having determined the luminosity $\lt$ sampled by the whole CCD 
(or IR array) frame, one can easily derive the average luminosity $L_{\rm
T}^{\rm pix}$ sampled by each pixel:
\begin{equation}
L_{\rm T}^{\rm pix}=\lt/N^{\rm pix},
\end{equation}
where  $N^{\rm pix}$ is the number of pixels in the detector.
The actual resolution element will generally exceed the size of one
pixel,
and in the case of ground-based observations it will depend on seeing.
The luminosity sampled by one resolution element is then 
 the product of $L_{\rm T}^{\rm pix}$ times the number of
pixels in each resolution element. In the sequel the term pixel will
be used for short, but everything said for pixels would apply 
equally well to the actual resolution elements.

Crowding has two main effects on the results of automatic reductions
of CCD frames by photometric packages. One obvious effect is the
reduced photometric accuracy: the fit to the stellar PSF is degraded
by the presence of neighboring stars, and the error in magnitudes
exceeds that expected from the pure Poisson photon noise. The less 
obvious effect is that occasionally two (or more) stars fall on
the same pixel, and the package  is not able to realize 
the multiple nature of the object which is then  mistaken as a single
star. The former effect results in broadened color-magnitude diagrams
and smeared luminosity functions, and one can easily learn how to
survive with such reduced photometric accuracy. The latter effect is 
instead much more
insidious, as it generates false stars that do not have a counterpart
in the real population, but may resemble objects with attractive
characteristics one may be wishing to find.

Let $\nj=\bt L_{\rm T}^{\rm pix}\tj$ be the number of stars in phase
``j'' per pixel; for $\nj <1$ this is also close to the probability
that a pixel contains one star in phase ``j''. Hence, the probability 
that a pixel contains two such stars is $\sim N_{\rm j}^2$, and the
number $N_{2j}$
of such events in the whole frame is:
\begin{equation}
N_{2j}=N_{\rm j}^2N^{\rm pix}=[\bt L_{\rm T}^{\rm pix}\tj]^2N^{\rm
pix}= [\bt\lt\tj]^2/N^{\rm pix}.
\end{equation}
Since $\lt$ is proportional to the surface brightness sampled by the
camera, one sees that the number of 2 j-star blends is proportional to
the square of the surface brigthness  expressed in physical units 
($\lsun/\squapp$). This {\it square law} can be easily
generalized to any kind of star pairs (e.g. the blend of a WD and a MS
star),
with the number $N_{\rm jk}$ of such ``jk'' blends in the frame being:

\begin{equation}
N_{\rm jk}=\nj N_{\rm k}N^{\rm pix}=[\bt\lt]^2\tj t_{\rm k}/N^{\rm pix},
\end{equation}
where the second equality applies only to PMS pairs. Hence, the number
of any kind of two star blends in the frame goes with the square of
the surface brightness of the target.
Moreover, the number of triplets 
goes with the cube of the surface brightness, that of quartets with
the fourth power, and so on.

This square law uncovers yet another insidious effect of crowding. 
The surface brightness of stellar systems can be very  high at the
center, then falling rapidly towards to outer regions. Therefore,
the number of star pairs sharing the same pixel will increase towards
the center as the square of the surface brightness, i.e. the
``exceptional'' stars will appear much more centrally concentrated than
the underlying light distribution. One may be tempted to conclude e.g., that 
the exceptional stars were produced by a recent burst of star
formation that took place in the central regions. 
A plausible hypothesis that may become a wishful finding.
From equation (9) it follows that  the number $N_{\rm jk}$ of jk blends
is also proportional to the square power of the actual resolution.

A safe criterion for accurate photometry states that meaningful
magnitudes can be obtained only for stars much brighter than $L_{\rm
T}^{\rm pix}$, i.e., for $L\gg L_{\rm
T}^{\rm pix}$. By contrast, there is no way to extract meaningful
information on individual stars fainter than this limit, i.e. for
 $L\lsim L_{\rm T}^{\rm pix}$. When this applies even
to the brightest stars in the population one enters the domain of the 
surface brightness fluctuation method (Tonry \& Schneider 1988), and
one had
better leave photometric packages aside: the frame is {\it burned out} almost
like an overexposed photographic plate.

Of course, the quality of the photometry degrades in a continuous
fashion
as the $L_{\rm T}^{\rm pix}$ limit is approached. For example, 
suppose that in a given frame the photometry of the brightest stars is
on the safe side of the criterion. Going to fainter objects photometry
will become more and more hazarduous as their luminosity approaches
 $ L_{\rm T}^{\rm pix}$. Along with real individual stars a photometric
package will deliver an increasing number of ``jk'' pairs in the same
magnitude interval. When working on such a mined terrain,
equation (9)  can be used to estimate the number of contaminants, and
it would not be wise to trust the output of a photometric package when
the estimated number of contaminants in a given magnitude bin becomes
a sizable fraction of the total star counts in the bin. 

This criterion for safe photometry can be made even more effective if
assisted by appropriate simulations and numerical experimets. For
example, comparing the results of running a given photometric package
on a full resolution frame, and then on the same frame with the
resolution having being artificially degraded (e.g. DePoy et
al. 1993).  Indeed, the ability to resolve blends into individual
components is code dependent, while (photon) S/N will also affect
the output.  In essence, a modest effort invested in such simulations
would {\it calibrate} the criterion, allowing it to predict with
greater accuracy the performance of a specific photometric package.

\section{Some Examples}

In this section a few concrete examples are presented to illustrate
the use and effectivness of the tools described in the previous
sections. These examples include the extreme cases one is interested
in at either the top or the bottom end of the luminosity function, and
involve both the frame sampling and the pixel sampling criteria.
These examples are restricted to the case of an old stellar
population, but the tools provided in this paper should allow the
concerned reader to deal with populations of any age distribution.

\subsection{Brightest Stars in the Bulge and Satellites of M31}

The red giant (RGB+AGB) luminosity function in nearby galactic
spheroids can in principle be used to infer whether an intermediate
age population in present in such objects, meaning with that a
population significantly younger than galactic globular clusters.
However, the presence of objects apparently brighter than the RGB tip
($\mbol\simeq -3.8$ for [Fe/H]$\simeq 0$) cannot be interpreted as
evidence for an intermediate age population, unless their number is 
significantly in excess of that of LPVs found in old, near solar
metallicity globulars, where LPVs extend up to
$\mbol\simeq -5.0$. A trace population of AGB stars even brighter than
this limit is represented by the progeny of blue stragglers, a minority
component in old stellar system. Finally,
under extreme crowding conditions apparently bright objects may
actually be the result of unresolved blends. 

 Table 2 reports sampling information for
various location in the bulge of M31, and in M32 and NGC 147, assuming
a common distance modulus of 24.4 mag. The sampled blue and bolometric
luminosities per \squapp\ ($\lb$ and $\lt$) take into account the
extinction correction assuming $E(B-V)=0.18$ (hence $A_{\rm B}=0.756$,
Han et al. 1997), and adopting $\lt=2.5\lb$. The extinction correction
accounts for the factor of 2 difference in some of the numbers with
respect to Table 1 in Renzini (1993).  The last column in Table 2
gives the number of pixels containing 2 RGB stars within 1 mag from
the RGB tip, for specific values of the size of the resolution element
(see below). The majority of these
events will produce artifact ``stars" brighter than the RGB tip, that
might be mistaken for bright AGB stars.

The first M31 line in Table 2 refers to the central $4''$ of M31, an
area that was observed by Davidge et al. (1997) with an adaptive
optics device with a pixel size of $0''.0343$ that -- using one of the
two M31 nuclei as a reference star -- allows a  resolution of
$0''.14$ (FWHM). The surface brightness in the field drops from
$\mu_{\rm B}\simeq 13.8$ mag/\squapp at the center, to $\sim 16.7$ at
the edge of the observed field (Schweizer 1979). Taking into account
the square-law effect (Section 3.2), a surface brightness $\mu_{\rm
B}\simeq 15$ mag/\squapp can be taken as representative of the filed.
Thus, each pixel collects $\sim 4500\,\lsun$ and
each reolution element $\sim 7.5\times 10^4\lsun$, comparable to that of
a typical globular cluster. The actual population of each resolution
element can be obtained scaling from Table 1, multiplying the last
column by 0.75. In particular, each resolution element contains $\sim
0.4$ LPVs and
$\sim 6.5$ RGB tip stars. No meaningful stellar photometry can be
extracted from such a frame, and the apparent ``bright AGB stars''
reported by Davidge et al.  must merely be peaks caused by the stocastic
fluctuation in the number of LPV and RGB stars per resolution element.
HST images of this central field could be used both to get an accurate
surface brightness distribution, and to attempt resolving into
individual components at least some of the apparently brightest objects.

\placetable{tbl2}

Moving to less crowded fields, let us suppose we observe at $2'$ from
the center of M31 with the NIC2 camera on HST. With a pixel size of
$0''.06$
and a field of view of $19''.2\times 19''.2$, NIC2 would sample a
total
luminosity $\sim 1.7\times 10^7\lsun$, or $\sim 160\,\lsun$ per pixel.
Scaling from Table 1, one infers that the NIC2 frame should include
$\sim 80$ LPVs. From equation (8) one gets that $\sim 20$ two-RGT
star events in one pixel should also be present in the frame
(cf. Table 2). 
As also reported in Table 2, one sees that at $4'$ from the center the
number of two-RGT blends has decreased a factor of ten, while the
number of LPVs decreases by a factor of 3, an illustration of the
`square law' mentioned in Section 3.2.
When taking into
account the  actual resolution will be a few times worse than one pixel,
one can conclude that NIC2 will barely be able to obtain meaningful
stellar photometry in M31 closer than $\sim 2'$ from the center.

The $2'$ and $4'$  M31 fields were also observed by Rich \& Mould (1991) and
Rich, Mould, \& Graham (1993) with a $\sim 1\,\squapp$ resolution. One
expects that in both fields two-RGT blends are more numerous than LPVs
(Renzini 1993), see also DePoy et al (1993) for detailed simulations. 
A better resolution is clearly required, and Rich \& Mighell (1995)
used pre-Costar HST Wide Field Camera data in an attempt to improve
upon previous ground based observations. The sharp core and extended
wings of the aberrated PSF forced them to abandon the PSF fitting
technique in the very crowded central and $2'$ fields they observed,
and taylor to the nature of the data an aperture photometry approach. 
In the central
field their $0''.4$ aperture samples an average $\sim 10^4\lsun$,
hence photometry of stars of comparable luminosity must be very uncertain.
Indeed, Rich \& Mighell admit that they ``cannot be {\it certain} that
stars brighter
than $I=19.5$ [corresponding to $\mbol=-4.5$] are intrinsically
luminous or accidentally measured bright due to photometric errors''.
The rather blue $R-I$ color of the detected objects also favors the bright
events being blends of RGB (or LPV+RGB) stars, rather than extended
AGB objects that would be rather red. In conclusion, NIC2 offers
our best hope to check whether or not the bulge of M31 harbors an
extended AGB made of intermadiate age stars, and yet this will be
possible only beyond some arcminutes from the center. 

Several studies have focussed on M32, the dwarf elliptical satellite
of M31. Table 2 shows that the two innermost fields are too crowded
for reliable stellar photometry with the resolution available to
Davidge \& Nieto (1992), each resolution element being $\sim
0.16\,\squapp$.  The situation was far better for the outer field
located at
$2'$ from the center, which was observed from the ground by Freedman
(1992) with
$\sim 0.3\, \squapp$ resolution elements.  With the average surface brightness
quoted by Freedman ($\mu_{\rm V}=21$, or
$\mu_{\rm B}=21.9$) over a field of view of 4000 $\,$ \squapp, one
expects $\sim 130$ LPVs and $\sim (0.3\times 0.7)^2\times
4000/0.3\simeq 600$ two-RGT blends in the frame. By contrast, Freedman
detected only $\sim 100$ stars brighter than the RGB tip.  However,
the {\it average} surface brightness of this field may actually be
significantly lower than mentioned by Freedman. From Peletier (1993) the
much lower {\it local} value $\mu_{\rm B}=24.6$ can be inferred at the $2'$
position (Renzini 1993), a value also reported in Table 2.  Assuming this
local value to be representative of the average over the actual field
of view, one obtains a $\sim 10$ times lower value of $N_{\rm LPV}$ and a
$\sim 100$ times lower value of $N_{\rm 2RGT}$. The apparent discrepancy
is at least in part a result of the strong $\mu_{\rm B}$-gradient
across the field, hence to estimate of the number of blends
one should take into account the actual surface brightness distribution
within the $\sim 1'$ field of view. It appears that the issue has been
resolved by recent observations with HST/WFPC2 at $1'-2'$ from the
center of M32, which did not reveal the presence of AGB stars so
(bolometrically) bright to be indicative of an intermediate age
component (Grillmair et al. 1996). Grillmair et al. conclude that
the optically bright objects seen in previous ground-based work 
were artifacts of crowding.

HST observations of another satellite of M31, NGC 147, have revealed
the presence of a population of stars up to $\sim 1.$ mag brighter in
$I$ than the RGB tip (Han et al. 1997). The Han et al. inner field is
centered on the nucleus of NGC 147, and extends to $\sim 100''$ from
the center along the major axis. 
From Table 2 one sees that at the very center of NGC 147 the surface
brightness is still low enough for accurate photometry to be done to
rather faint luminosities. With an $0''.1$ pixel size, each WFPC2
pixel will sample $\sim 43\lsun$, and serious crowding problems will
be encountered more or less at the level of the HB. Therefore, image
blends
can be excluded as the origin of the extended AGB in this galaxy.

The NGC 147 integrated magnitude within the $r=100''$ isophote 
is $B_\circ \simeq 10.6$ (Hodge 1976), and the total blue
luminosity sampled by WFPC2 on the inner field is roughly $\lb\simeq
5\times 10^7\lsun$, or $\lt\simeq 1.25\times 10^8\lsun$. This may
somewhat overestimate the actual luminosity sampled within the
stealthy perimeter of the WFPC2 field of view, 
so $\sim 10^8\lsun$ seems to be a more
conservative estimate. (The luminosity actually sampled by WFPC2 could
be obtained as described in Section 3.1.)  Scaling from Table 1 one
then expects $\sim 500\pm250$ LPVs to lie above the RGB tip, while Han
et al find $\sim 250$ such stars.  The large statistical uncertainty
comes  from the estimated LPV lifetime being based on 4 ($\pm 2$)
LPVs in 47 Tuc.  The average metallicity of NGC 147 ([Fe/H]=--0.9) is
close to that of 47 Tuc (--0.7), yet due to the large metallicity
dispersion only a fraction of the stellar population is expected to
exceed the metallicity threshold ([Fe/H]$\gsim -1$) above which
galactic globular clusters harbour LPVs (Frogel \& Elias 1988).  
Finally, scaling from Table 1 one expects to find $\sim 75$ TP-AGB
stars being the progeny of BSs, hence reaching brighter luminosities
on the AGB than the bulk population of single stars.

In
conclusion, a population of LPVs similar to the ones in old, metal
rich globulars such as 47 Tuc, appears able to account for the
extended AGB of NGC 147 without appealing to an intermediate age
population.  The gradient in the frequency of bright AGB stars can
well result from the higher proportion of metal rich stars near the
center that was found by Han et al., rather than from a recent burst of
star formation. The above discussion has also indicated that crowding
has so far seriously limited the photometry of individual stars in the inner
regions of M32 and the bulge of M31, while high resolution HST imaging
of an outer
field in M32 has failed to detect an extended AGB. From all this it
appears that no compelling evidence has yet emerged for the presence
of a major intermediate age component in any of these systems.

\subsection{The Main Sequence Turnoff of the M31 Bulge}
As well known, the best way of determining the age of a stellar
population, or to detect an unequivocal age spread, is to access
directly the main sequence turnoff. Experience has shown that
photometry should reach at least two magnitudes below the turnoff with
$S/N\gsim 10$ in order to determine the turnoff magnitude with
adequate accuracy (say, $\sim 0.1$ mag, corresponding to a $\sim 10\%$
error in age).

With $M_{\rm V}^{\rm TO}\simeq 5$, as
appropriate for a metal rich population 15 Gyr old, the turnoff is 
to be found at $V\simeq 29.4$, and therefore one should reach $V\simeq 31.4$
with the appropriate $S/N$. This seems to be out of reach even for an
heroic effort with HST. NGST instead will have a collecting area and QE 
adequate to reach this limit, but let us consider the crowding
problem.

For this example let us focus on the outermost of the M31 fields in
Table 2, which corresponds to a projected distance of $\sim 2$ kpc
from the center, in fact near the edge of the bulge.  
With an absolute bolometric surface brightness of $\sim
2400\, \lsun/\squapp$, each $0''.06$ pixel of NGST/NIR camera
(Stockman 1997) will sample $\sim 8.6\lsun$. Two magnitudes below
turnoff corresponds to stars of $\sim 0.7\msun$, assuming the turnoff
itself is at $0.9\msun$.  Hence, integrating the IMF following the
equation (3) prescription, one gets that on average each pixel will
contain $\sim 3.6$ stars brighter than $V=31.4$. It appears
that the main sequence turnoff of the M31 bulge will be out of reach
due to crowing even with NGST.
A resolution some ten times better (i.e., $\sim 0''.005$) would be
needed,
something that would require a baseline of $\sim 80$m, either with an
interferometer, or with a single dish.

\subsection{The Bottom of the Main Sequence in the Galactic Bulge}
The determination of the intial mass function (IMF) of the galactic
bulge is of great astrophysical interest. It
represents a unique chance of measuring directly the lower IMF of an
old, metal rich population that may have formed in a major starburst
some
15 Gyr ago. Moreover, knowing the IMF of the bulge would be of great
help in interpreting the results of the various microlensing
experiments now under way.

This example focusses on a bulge field $\sim 6^\circ$ south of the
galactic center. With a surface brightness $\mu_{\rm V,\circ}=19.7$ mag/\squapp
(corrected for extinction) and a true distance modulus 14.5 mag, the
absolute surface brightness is   
$M_{\rm
V,\circ}=5.2$ mag/\squapp, or 1.5 bolometric $\lsun/\squapp$, using
$\lt=2.1\lv$. Among currently available telescopes and cameras, NICMOS
on  HST offers the best chance of reaching the bottom of the
main sequence. With a FoV of $19''.2$, the NIC2 camera will sample $\sim
420\lsun$, and following  equation (6) the frame will include $\sim
8000$ stars. With $256^2$ pixels, there will be on average $\sim 8$
pixels available for each star, and it may turn out to be very difficult to
do accurate photometry down to the hydrogen burning limit. However,
this  assumes a Salpeter's IMF slope ($x=1.35$) all the way
down to $\sim 0.1\msun$. If the IMF of the bulge flattens out
below $0.6 \msun$, as it does in the solar neighborhood (Gould, Bahcall
\& Flynn 1997), then one expects to find $\sim 10$ times fewer 
stars in the frame, and there will be over 100 pixels available 
for each stellar
image. This seems to be perfectly adequate for accurate photometry
down to the MS limit, provided the integration time is long enough.
However, given the strong surface brightness gradient, it appears
difficult to obtain complete IMF information for fields somewhat
closer to the galactic center.

\section{Conclusions}

Simple tools have been presented that may help the astronomer wishing
to extract as much
sound science as possible from CCD/IR array data on stellar populations.
The main conclusions of this paper can be summarized as follows.

(1) While planning the CCD/IR array observations of stellar populations in 
globular clusters, in the Galactic Bulge, or in nearby resolvable
galaxies the total 
luminosity sampled by the frame (in $\lsun$ units) should be estimated from
the known surface brightness distribution of the target objects.

(2) On the basis of this ``frame luminosity", the luminosity sampled by each
pixel of the camera (or by each expected resolution element) should also
be evaluated.

(3) On the basis of the frame luminosity the astronomer can estimate
with very good accuracy the number of stars in each evolutionary phase that
will be framed by the camera, thus allowing one to decide whether the
sampling is statistically adequate for the specific evolutionary phases under
investigation.

(4) On the basis of the ``pixel" luminosity the astronomer can estimate with
very good accuracy down to which magnitude one can trust the result of 
individual star photometry as performed by current photometric packages.

(5) Frame sampling and pixel sampling set conflicting requirements, the former
asking for regions of high surface brightness to be observed in order to secure
a statistically adequate sample of stars, the latter instead asking for regions
of low surface brightness to be observed in order to secure reliable
photometry of individual stars. The tools provided in this paper allow the 
astronomer to proceed very rapidly with the necessary optimization, or
realize that a certain scientific goal is not reachable with the
available technology.

(6) After the data have been taken, the frame luminosity should be
estimated directly from the frame itself, after the frame has been
properly dark, bias, and sky subtracted and calibrated. This will
provide the most accurate estimate of both the frame and pixel
luminosity as possible.

(7) A basic criterion for the limit imposed by crowding states that
reliable photometry can be obtained only for those stars that are brighter than
the average luminosity sampled by each pixel (resolution element).
Experiments that are easy to imagine should allow one to more finely
calibrate  this criterion, establishing in a quantitative way -- and
for any specific photometric package -- how the
photometry of individual stars degrades as this limit is approached, and
to which extent star counts in a given magnitude bin are contaminated by 
blends of stars in fainter magnitude bins.

(8) Some straightforward checks should be made when photometric
packages restitute exceptionally bright objects, in order to ascertain
whether they are real stars rather than the result of accidental
blends of fainter stars sharing the same resolution
element. This second option should be carefully evaluated especially when 
the surface density of such objects appears to be rougly proportional to the 
square power of the surface brightness.
 
(9) The {\it artificial star} experiments as currently done by standard
photometric packages may be adequate to estimate the completeness of star
counts as a function of magnitude. However, they
would require to repeat the experiment nearly as many times as the
number of 
stars in the frame in order to assess the extent of the {\it migration}
of a fraction of stars towards brighter magnitudes  due to blending with
other stars.

(10) An application of these tools to existing photometric data for
the bulge of M31, M32 and NGC 147 indicates that no solid evidence has
yet emerged for the presence of an intermediate age population in
these objects.

I wish to thank Mike Rich for extensive discussions -- over the last
several
years -- on the effects of crowding 
on the results of package photometry of Local Group galaxies.
I wish also to thank Russell Cannon for a critical reading of the
manuscript which resulted in an improved presentation. I am grateful
to Claudia Maraston for her help in producing Figure 2.

\clearpage

\begin{deluxetable}{lrrrr}
\footnotesize
\tablecaption{Star Numbers for $\lt = 10^{5} \lsun$,
Age = 15 Gyr, and $Z=Z_\odot$ \label{tbl1}}
\tablewidth{0pt}
\tablehead{
\colhead{Evolutionary Phase} & \colhead{$t_{j}$ (yr)}   & \colhead{$N_{j}$}}
\startdata
MS  &  $>10^{10}$        & $1.9\times 10^6$      \\
SGB &  $3\times 10^9$    & 6000  \\
RGB &  $6\times 10^8$            & 1200  \\
RGBT&  $5\times 10^6$    & 10    \\
HB  &  $10^8$            & 200   \\
E-AGB  & $1.5\times 10^7$  & 30    \\
TP-AGB  & $10^6$  & 2    \\
LPV  & $2.5\times 10^5$  & 0.5   \\
P-AGB& $3\times 10^5$            & 0.6   \\
PN   &  $2.5\times 10^3$  & 0.005\\
WD   &  $10^9$            & 2000 \\
BS   &  $2\times 10^9$    & 200  \\
BS-TP-AGB& $10^6$            & 0.06 \\  
\enddata
\end{deluxetable}
\clearpage

\begin{deluxetable}{lrrrrrrrr}
\footnotesize
\tablecaption{Stellar Population Sampling in the
                        Bulge and Satellites of M31 \label{tbl2}}
\tablewidth{0pt}
\tablehead{
\colhead{$r$} & \colhead{$\mu_{\rm B}$}   & 
\colhead{$\lb$} & \colhead{$\lt$} & \colhead{$N_{\rm LPV}$} &
\colhead{$N_{\rm RGT}$} & \colhead{$N_{\rm 2RGT}$}\\
\colhead{} & \colhead{mag/\squapp} & \colhead{$\lsun$/\squapp} &
\colhead{$\lsun$/\squapp} & \colhead{stars/\squapp} &
\colhead{stars/\squapp} & \colhead{events}} 
\startdata
& & & M31 Bulge & & & \\

\tableline
$0''-4''$     &  15    &$1.5\times 10^6$&$3.8\times 10^6$   &20   &410 & $-$\\
2$^\prime$& 19.8\hfil &$1.8\times 10^4$ &$4.5\times 10^4$   &0.24  &4.9 & 21\\ 
4$^\prime$&   21\hfil  &$6.0\times 10^3$ &$1.5\times 10^4$    &0.06&1.6& 2 \\
$11'.5$   &   23       &$9.5\times 10^2$ &$2.4\times 10^3$    &0.01  &0.2& $-$\\

\tableline

& & & M32$\;\;$ & & & \\

\tableline
$5^{\prime\prime}$ &17\hfil &$2.4\times 10^5$ &$6.0\times 10^5$ &3 &66 & $-$\\
$27^{\prime\prime}$&18\hfil &$9.6\times 10^4$ &$2.3\times 10^5$ &1.2&23& $-$\\
$2^{\prime}$ &21.9\hfil    &$2.6\times 10^3$ &$6.5\times 10^3$ &0.03&0.7 &600\\
$2^{\prime}$ &24.6\hfil    &$2.2\times 10^2$ &$5.5\times 10^2$
&$0.003$&0.06& 6 \\
\tableline

& & & NGC 147$\,\,$ & & & \\

\tableline
$0''.0$& 22.55 & $1.4\times 10^3$ & $4.3\times 10^3$& 0.02 & 0.4 & $-$\\
\enddata
\end{deluxetable}
\clearpage

% Camera-ready tables, produced with either the apjpt4 or aj_pt4 style files,
% can be referenced within a table environment using \dummytable.  This acts
% like a place holder and bumps the table counter.   For this particular
% manuscript, tbl-3 refers to the table in file samp2tbl.tex.

% This is the last table for this paper (as well as the first), so we
% should follow it with a \clearpage.  In order to force all the floating
% tables out of their buffers and onto vertical page lists, we must use
% \clearpage rather than \newpage. 

\clearpage

\clearpage
\figcaption[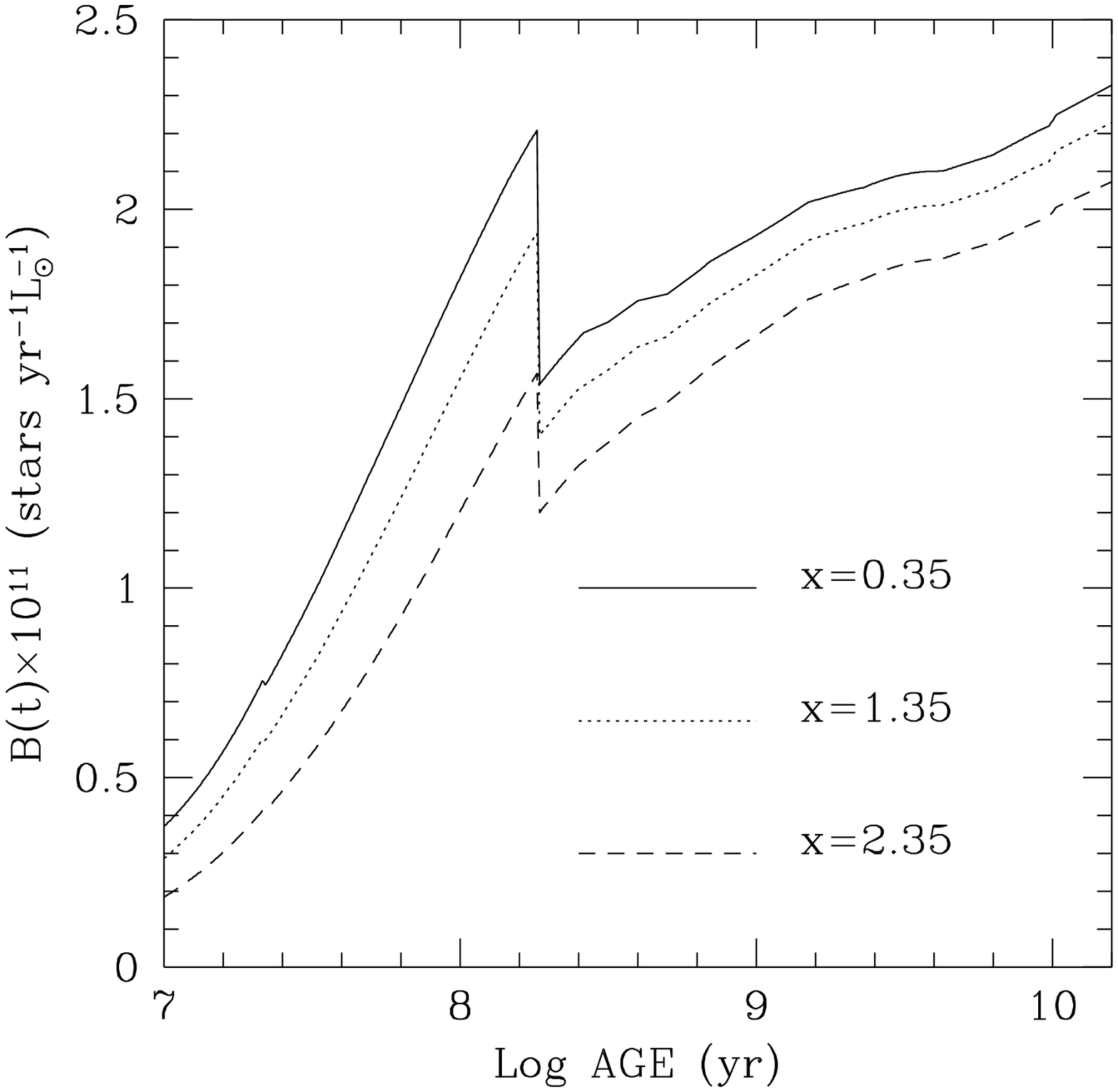]{The specific evolutionary flux, i.e. the number
of stars evolving off the main sequence per year and per unit (solar)
luminosity of the parent stellar population, for three values of the
slope of the initial mass function as indicated. The sharp
discontinuity at $\sim 10^8$ ye is caused by the appearence of an
extended TP-AGB phase.
\label{fig1} }

\figcaption[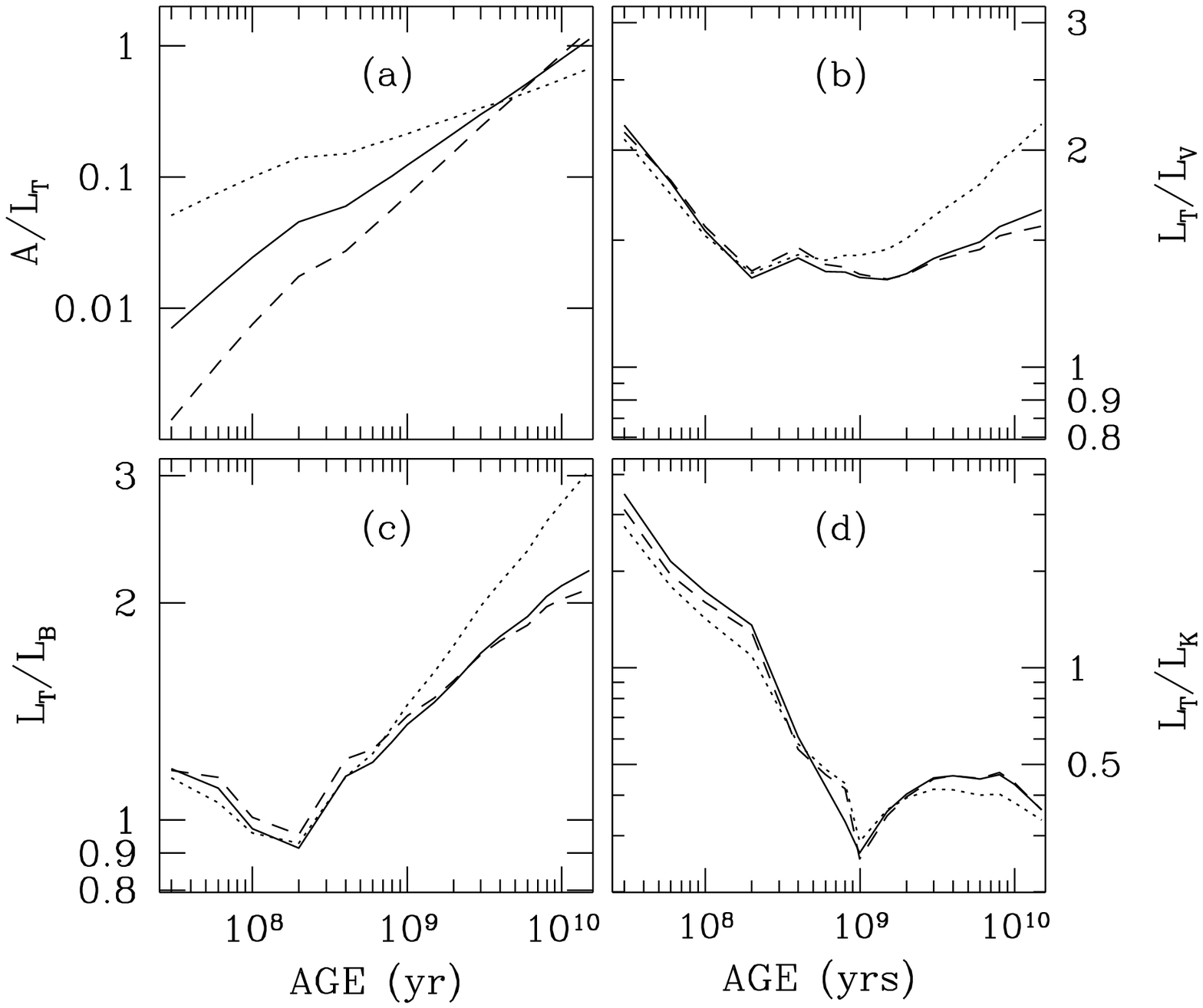]{(a) The ratio of the scale factor $A$ of the IMF
to
the total bolometric luminosity of a stellar population as a function
of its age. (b) The ratio of the bolometric luminosity to the $V$-band
luminosity as a function of age for a solar metallicity population.
(c) The same for the $B$-band luminosity. (d) The same for the
$K$-band luminosity. \label{fig2}}

\clearpage

\plotone{pixelfig1.ps}

\clearpage

\plotone{pixelfig2.ps}

\end{document}